\documentclass{jpp}
\usepackage{graphicx}

\usepackage[utf8]{inputenc}
\usepackage[T1]{fontenc}
\usepackage{amsmath}

\usepackage{dcolumn}
\usepackage{bm}
\usepackage{caption}
\usepackage{subcaption}
\usepackage{xcolor}
\usepackage{mathptmx}
\usepackage{etoolbox}
\usepackage{natbib}
\usepackage{amssymb}
\usepackage{hyperref}
\usepackage{tikz}
\usepackage{ulem}

\newcommand{\iotabar}{\iota\!\!\!\!\text{-}}
\newcommand{\ab}[1]{#1}

\shorttitle{Equilibrium $\beta$-limits dependence on bootstrap current in classical stellarators}
\shortauthor{A. Baillod, J. Loizu, Z. Qu, H. P. Arbez, J. P. Graves}

\title{Equilibrium $\beta$-limits dependence on bootstrap current in classical stellarators}

\author{A. Baillod\aff{1}
  \corresp{\email{antoine.baillod@epfl.ch}},
 J. Loizu\aff{1},
 Z. Qu\aff{2},
 H. P. Arbez\aff{1},
 \and J. P. Graves\aff{1}}

\affiliation{\aff{1}Ecole Polytechnique F\'ed\'erale de Lausanne, Swiss Plasma Center, CH-1015 Lausanne, Switzerland
\aff{2}School of Physical and Mathematical Sciences, Nanyang Technological University, 637371 Singapore, Singapore}

\begin{document}

\maketitle

\begin{abstract}
	While it is important to design stellarators with high magneto-hydrodynamic (MHD) stability $\beta$-limit, it is also crucial to ensure that good magnetic surfaces exist in a large range of $\beta$ values. As $\beta$ increases, pressure-driven currents perturb the vacuum magnetic field and often lead to the emergence of magnetic field line chaos, which can worsen the confinement and is the cause of another kind of $\beta$-limit, the so-called equilibrium $\beta$-limit. In this paper, we explore numerically the dependence of the equilibrium $\beta$-limit on the bootstrap current strength \ab{in a classical stellarator geometry} using the Stepped Pressure Equilibrium Code (SPEC). We develop a diagnostic to determine whether or not magnetic islands are expected to participate significantly to radial transport, and we build an analytical model to predict the expected equilibrium $\beta$-limit, which recovers the main features of the numerical results. This research opens the possibility to include additional targets in stellarator optimization functions, provides additional understanding on the existence of magnetic surfaces at large $\beta$, and is a step forward in the understanding of the equilibrium $\beta$-limit.
\end{abstract}

\section{\label{sec:intro}Introduction}
\normalem
In magnetic fusion devices such as stellarators, zeroth order confinement of particles and energy is obtained by constructing an equilibrium with magnetic surfaces.
Magnetic islands and magnetic field line chaos can be detrimental to confinement, \textit{i.e.} they can contribute to the radial transport of particle and energy \citep{Hudson2010}.
While it is possible to design equilibria with good magnetic surfaces in a vacuum \citep{Cary1985,Cary1986,pedersenConfirmationTopologyWendelstein2016}, pressure-driven plasma currents, such as diamagnetic, Pfirsch-Schl\"uter and bootstrap currents, perturb finite pressure equilibria, and, at a sufficiently large pressure, magnetic islands and chaos emerge.

A pressure increase can also sometimes heal magnetic islands \citep{Bhattacharjee1995}. While this mechanism can improve confinement locally, other islands might open elsewhere in the plasma as $\beta$ increases.
There is thus a critical value of $\beta$ at which magnetic islands open and magnetic field line chaos emerges. This defines an \emph{equilibrium $\beta$-limit}.
Note however that the equilibrium $\beta$-limit is a "soft" limit, since crossing it does not lead to a loss of control of the plasma. Additional input power may however leak through the damaged magnetic surfaces more easily \citep{rechesterElectronHeatTransport1978}, thereby preventing an increase of $\beta$. Crossing the equilibrium $\beta$-limit may thus not be as concerning as crossing a stability limit (which may lead to plasma disruptions), but it still limits the overall performance of the reactor. It is consequently of crucial importance to understand these equilibrium $\beta$-limits better, especially for the operation of existing experiments and the design of new machines. Configurations where good magnetic surfaces are preserved over a large range of $\beta$ have to be sought, which will help to ultimately identify configurations with large enough equilibrium $\beta$-limit.

In the case of a classical stellarator, \citet{Loizu2017} proposed a model for the equilibrium $\beta$-limit of a configuration with zero net toroidal current as well as one with a fixed edge rotational transform. 
Other studies computed high $\beta$ equilibria in a number of experimentally relevant stellarator configurations and predicted the emergence of magnetic field line chaos at sufficiently large $\beta$ - see, for example, the calculation by \citet{suzukiTheoreticalStudiesEquilibrium2020} in the Large Helical Device (LHD) and \citet{Reiman2007} in Wendelstein 7-AS (W7-AS). However, to the authors knowledge, no attempt has been made to analytically model the impact of the bootstrap current on the equilibrium $\beta$-limit, and to determine how this critical $\beta$ depends on the device parameters.

We propose to extend the work of \citet{Loizu2017} to the case of a classical stellarator with bootstrap current. We use the Stepped Pressure Equilibrium Code (SPEC) to compute a large number of free-boundary equilibria at different $\beta$, including the effect of bootstrap current. SPEC has been chosen for its speed, its capability to describe equilibria with magnetic islands and chaos, and the possibility to calculate free-boundary equilibria \citep{Hudson2020c} with a constrained toroidal current profile \citep{Baillod2021}.
SPEC has been verified in stellarator geometry \citep{Loizu2016}, and its core algorithm has been improved to run faster and to be more robust \citep{Qu2020}. It has been successfully applied to study current sheets at rational surfaces \citep{Loizu2015,Loizu2015a,huangNumericalStudyFunction2022}, ideal linear instabilities \citep{Kumar2021,Kumar2022}, tearing mode stability \citep{Loizu2019} and non linear saturation \citep{Loizu2020}, penetration of resonant magnetic perturbations in the ideal limit \citep{Loizu2016a} and relaxation phenomena in reversed field pinches \citep{Dennis2013a,Dennis2014,quSteppedPressureEquilibrium2020}.

To numerically identify the equilibrium $\beta$-limit, \citeauthor{Loizu2017} used a diagnostic based on the \emph{volume of chaos}, \textit{i.e.} the volume of plasma occupied by chaotic field lines, which were identified by measuring their fractal dimension \citep{Meiss1992c}. However, this approach is too pessimistic since some chaotic magnetic field lines might be able to preserve confinement \citep{Hudson2008}. An alternative approach, proposed by \citet{paulHeatConductionIrregular2022}, is to measure the \emph{effective volume of parallel diffusion}. This measures the fraction of plasma volume over which the local parallel transport dominates over the perpendicular one in setting the radial transport. Contrary to the volume of chaos,  this approach takes into account only sufficiently large resonances that do participate significantly to the radial transport. In this paper, we follow \citeauthor{paulHeatConductionIrregular2022} and measure the {equilibrium $\beta$-limit} by taking the $\beta$ above which the radial transport generated by damaged magnetic surfaces represents a significant fraction of the total radial transport.


This paper is organized as follows. In section~\ref{sec.spec}, the equations solved by SPEC are recalled. In section~\ref{sec.equil}, we construct free-boundary equilibria in a rotating ellipse geometry, and construct a bootstrap current model. In section~\ref{sec.diag}, a new diagnostic is developed to measure the equilibrium $\beta$-limit and compared to the volume of chaos. In section~\ref{sec.analyticalmodel}, we derive an analytical model to explain the numerically obtained equilibrium $\beta$-limit. Finally, some concluding remarks are provided in section~\ref{sec.conclusion}.

\section{The Stepped-Pressure Equilibrium Code}
\label{sec.spec}
\begin{figure}
	\centering
	\begin{tikzpicture}[scale=0.8]
		\node (fig) at (0,0) {
			\includegraphics[width=.8\linewidth]{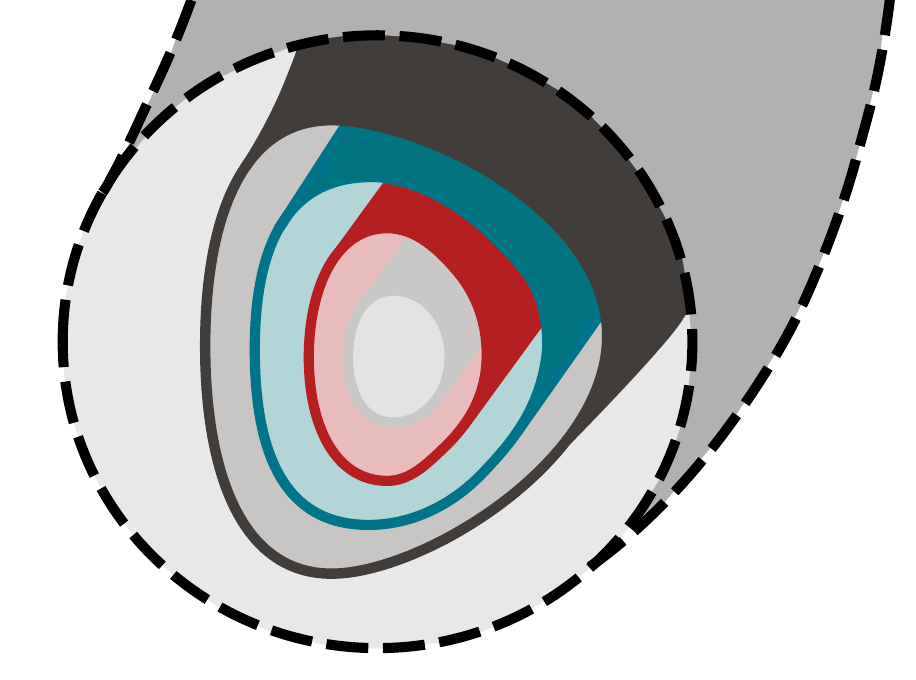}
		};
		\node (v1) at (-.9,-0.5) {$\mathcal{V}_1$};
		\node (v2) at (-1.35,-1.5) {$\mathcal{V}_2$};
		\node (v3) at (-1.8,-2.25) {$\mathcal{V}_3$};
		\node (v4) at (-2.4,-2.9) {$\mathcal{V}_4$};
		\node (I1) at (-0.2,0.7) {$\mathcal{I}_1$};
		\node (I1) at (.25,1.2) {\color{white}$\mathcal{I}_2$};
		\node (I1) at (0.75,1.8) {\color{white}$\mathcal{I}_3$};
		\node (I1) at (1.4,2.5) {\color{white}$\mathcal{I}_4=\Gamma_{PB}$};
		\node (cb) at (1.9,-4.1) {$\Gamma_{CB}$};
		\node[rotate = 40] (vac) at (1.1,-2.8) {Vacuum};
	\end{tikzpicture}
	\caption{Sketch of a SPEC equilibrium with four volumes. The plasma boundary, $\Gamma_{PB}=\mathcal{I}_4$, is the dark gray surface and the computational boundary, $\Gamma_{CB}$, is the light gray surface.}
	\label{fig.spec volumes}
\end{figure}
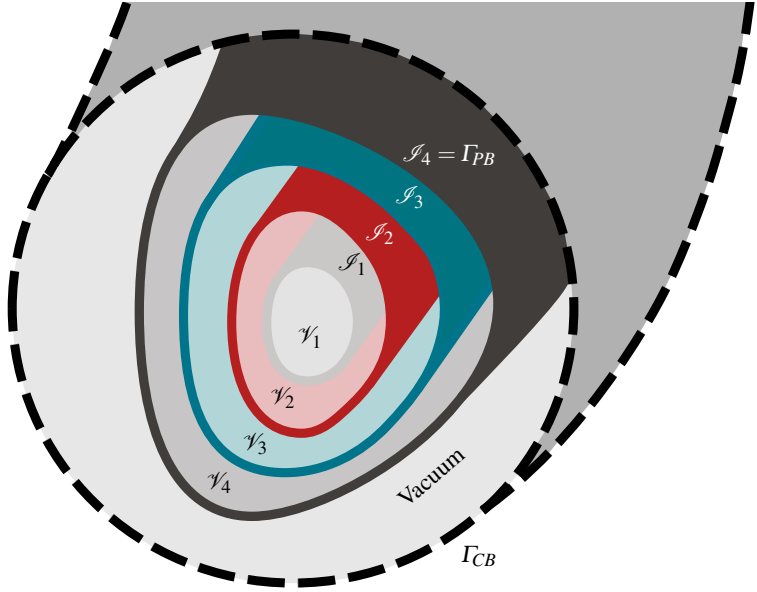

SPEC finds three-dimensional (3D) free-boundary magneto-hydrodynamic equilibria with stepped-pressure profiles.
Pressure steps are supported by a finite number of nested toroidal surfaces $\mathcal{I}_l$, thereby defining $N_{vol}$ nested volumes $\mathcal{V}_l$ with constant pressure $p_l$, with $l\in\{1,\allowbreak\ldots,\allowbreak N_{vol}\}$ (see Figure~\ref{fig.spec volumes}).

The magnetic field $\mathbf{B}$ in each volume is allowed to reconnect and can form magnetic islands and chaotic field lines, while the volume interfaces are constrained to be nested magnetic surfaces.
The magnetic field in each volume is a force-free field described by a Taylor state \citep{Taylor1974,Taylor1986},
\begin{equation}
	\nabla\times\mathbf{B} = \mu_l\mathbf{B}, \label{eq.taylor state}
\end{equation}
with $\mu_l$ a constant specific to the volume $\mathcal{V}_l$, and the solution to Eq.(\ref{eq.taylor state}) depends on the geometry of the interfaces enclosing the volume $\mathcal{V}_l$. SPEC finds the geometries of interfaces $\mathcal{I}_l$ such that force balance is satisfied,
\begin{equation}
	\left[\left[p+\frac{B^2}{2\mu_0}\right]\right]_l=0, \label{eq. force balance}
\end{equation}
where $\mu_0$ is the vacuum permeability, $p$ is the pressure and $[[\cdot]]_l$ denotes the jump across the interface $\mathcal{I}_l$. Equation (\ref{eq. force balance}) is the local equivalent to the more common force-balance condition $\mathbf{j}\times\mathbf{B}=\nabla p$.

The last interface defines the plasma boundary $\Gamma_{PB}$. The plasma is surrounded by a vacuum region (where $p_l=0$ and $\mu_l=0$), itself bounded by a computational boundary $\Gamma_{CB}$ that lies outside the plasma and inside the coils. The toroidal surface $\Gamma_{CB}$ is an otherwise arbitrary mathematical surface and not necessarily a magnetic surface, \textit{i.e.} generally $\mathbf{B}\cdot\mathbf{n}\neq 0$ on $\Gamma_{CB}$, with $\mathbf{n}$ a vector normal to $\Gamma_{CB}$. Note that the plasma averaged $\beta$ is evaluated from a SPEC equilibrium as
\begin{equation}
	\beta = \frac{1}{V}\sum_{l=1}^{N_{vol}}2\mu_0 p_l\iiint_{\mathcal{V}_l} \frac{dv}{B^2} ,
\end{equation}
with $V$ the volume enclosed by $\Gamma_{PB}$.

Free-boundary equilibria are determined by providing the total current \ab{flowing through the torus hole}, $I_c$, the geometry of the computational boundary, and the harmonics of the vacuum field (produced by the coils) normal to the computational boundary.
In addition, SPEC requires as input, in each volume, the enclosed toroidal flux $\psi_{t,l}$, the pressure $p_l$, the net toroidal current within the volume $I^v_{\phi,l}$ (closely related to the constant $\mu_l$), and the net toroidal current flowing at each interface $I^s_{\phi,l}$, which is a surface current.

Volume currents $I^v_{\phi,l}$ represent all externally driven currents, such as ohmic current, Electron Cyclotron Current Drive (ECCD) or Neutral Beam Current Drive (NBCD). Surface currents $I^s_{\phi,l}$ are all pressure-driven currents, such as diamagnetic, Pfirsch-Schl\"uter or bootstrap current, or island shielding currents.
Further details about the SPEC algorithm and implementation can be found in \citep{Hudson2012,Hudson2020c,Baillod2021}.

\section{Rotating ellipse with bootstrap current} \label{sec.equil}

We study the case of a rotating ellipse (sometimes also called classical stellarator) with an analytical bootstrap current model.
While a rotating ellipse is arguably a simple geometry, it is still relevant since all stellarators without \ab{magnetic axis} torsion are rotating ellipses close to the magnetic axis \citep{helanderTheoryPlasmaConfinement2014}. An experimental instance of rotating ellipse was, for example, the Wendelstein VII-A stellarator \citep{Grieger1985}.

\begin{figure}
	\centering
	\includegraphics[width=.7\linewidth]{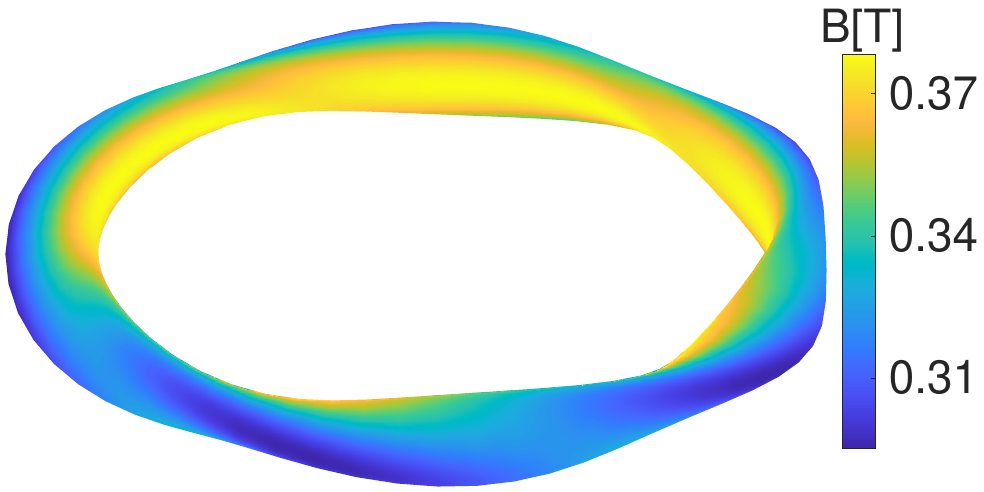}
	\caption{Computational boundary described by Eqs.(\ref{eq.cb_r})-(\ref{eq.cb_z}). Colors indicate the magnetic field strength in vacuum.}
	\label{fig. modB boundary}
\end{figure}

We choose a computational boundary $\Gamma_{CB}$  (see Figure~\ref{fig. modB boundary}) using standard cylindrical coordinates $\mathbf{x}=R_{CB}(\theta,\phi)\mathbf{\hat{e}}_R +Z_{CB}(\theta,\phi)\mathbf{\hat{e}}_Z$, with
\begin{align}
	R_{CB}(\theta,\phi) &= R_0 + R_{10}\cos(\theta) + R_{11}\cos(\theta-N_{fp}\phi)\label{eq.cb_r}\\
	Z_{CB}(\theta,\phi) &= Z_{10}\sin(\theta) + Z_{11}\sin(\theta-N_{fp}\phi)\label{eq.cb_z},
\end{align}
with $N_{fp}=5$ the number of field periods, $R_0=10$m, $R_{10}=-Z_{10}=1$m, $R_{11}=Z_{11}=0.25$m. The effective minor radius is $a_{\text{eff}}=\sqrt{r_{min}r_{max}}$ with $r_{min}=R_{10}-R_{11}$ and $r_{max}=R_{10}+R_{11}$ the minor and major radii of the ellipse respectively. We define $\epsilon_a=a_{\text{eff}}/R_0$ as the inverse aspect ratio \ab{at the plasma boundary}.

\ab{We assume that a coil system exists such that $\mathbf{B}_c\cdot\mathbf{n}=B_v\hat{\mathbf{e}}_z\cdot\mathbf{n}$ on $\Gamma_{CB}$, where $\mathbf{B}_c$ is the magnetic field produced by the coils, and $B_v\hat{\mathbf{e}}_z$ is a constant vertical field, which is applied to keep the plasma within the computational boundary at high $\beta$. We set $B_v=-0.03$T. This vertical field has little to no impact on the results presented hereafter; its only purpose is to keep the plasma within the volume defined by $\Gamma_{CB}$.  We fix the total current flowing in the torus hole to $I_c=17.1$MA, which determines the toroidal flux enclosed by the computational boundary, $\psi_{t,V}=1\text{Tm}^2$}.


We choose a pressure profile with a linear dependence on the toroidal flux, \textit{i.e.} $p=p_0(1-\psi_t/\psi_a)$, with $p_0$ a free parameter and $\psi_a=0.25\text{Tm}^2$ the total toroidal flux enclosed by the plasma boundary $\Gamma_{PB}$. We approximate the pressure profile with seven steps of equal magnitude $[[p]]_l=p_0/N_{vol}$. We thus define seven plasma regions, \textit{i.e.} $N_{vol}=7$, surrounded by a vacuum region. This means that $\psi_{t,l}=(l-1)\psi_a / N_{vol}$ and $p_l=p(\psi_{t,l})$. The number of volumes determines how the pressure profile is represented --- more volumes means more and smaller pressure steps. As each interface is a discrete constraint on the magnetic topology, increasing the number of volumes reduces the available space for reconnection and thus the maximum size of magnetic islands and regions of magnetic field line chaos.  In this paper, we are however interested in the onset of loss of magnetic surfaces, which is not affected by the volume available for islands to grow. Therefore our results only depends weakly on the number of volumes \ab{(see appendix \ref{app.nvol dep})}.

Two current profiles have to be provided to SPEC: the profile of volume currents, $\{I^v_{\phi,l}\}$, and the profile of surface currents $\{I^s_{\phi,l}\}$ (see section \ref{sec.spec}). Here we study the case of an equilibrium with zero externally driven currents and with bootstrap current. No externally driven currents implies, in SPEC, that there are no currents in the plasma volumes, \textit{i.e.}
\begin{equation}
	I^v_{\phi,l} = 0.\label{eq.volume current}
\end{equation}
	
The bootstrap current is a pressure-driven current, and is consequently described by a surface current at the volume's interfaces. We model it with
\begin{equation}
	I^s_{\phi,l} = -C \left(\frac{\psi_{t,l}}{\psi_a}\right)^{1/4} \left[\left[p\right]\right]_l, \label{eq.bootstrapmodel}
\end{equation}
where $(\psi_t / \psi_a)^{1/4}\approx\ab{\sqrt{\epsilon/\epsilon_a}}$ is related to the fraction of trapped particles, with $\epsilon$ the inverse aspect ratio; $[[p]]_l$ is a measure of the local pressure gradient; and $C$ is a coupling constant, in $[APa^{-1}]$, which controls the strength of the bootstrap current in the system. 
A full neoclassical calculation of the bootstrap current, for example with the SFINCS code \citep{landremanComparisonParticleTrajectories2014}, would require the density and temperature profiles as inputs --- and the freedom in the choice of the coupling constant $C$ reflects the freedom in these profiles.

The current density associated to the current in Eq.(\ref{eq.bootstrapmodel}) is
\begin{equation}
	j_{\phi,l} = -\frac{C\psi_a}{\pi a_{\text{eff}}^2}\left(\frac{\psi_{t,l}}{\psi_a}\right)^{1/4}\frac{dp}{d\psi_t}. \label{eq.current density continuous}
\end{equation}	
Note that if 
\begin{equation}
	C = C_0 \equiv \frac{\sqrt{\epsilon_a}R_0}{\iotabar_v B_0}, \label{eq. def C0}
\end{equation}
with $\iotabar_v$ the edge rotational transform in vacuum and $B_0$ such that $\mu_0I_c= 2\pi R_0 B_0$, Eq.(\ref{eq.current density continuous}) reduces to the well-known large-aspect ratio tokamak bootstrap current approximation \citep{helanderCollisionalTransportMagnetized2002},
	
\begin{equation}
	j_{\phi} = \sqrt{\epsilon_a}R_0\frac{d p}{d\psi_p},
\end{equation}
where $\psi_p$ is the poloidal flux, and we made the approximation $d\psi_p/d\psi_t=\iotabar\approx\iotabar_v$. We normalize $C$ by $C_0$, and define $\hat{C}\equiv C / C_0$. In the case of a large aspect ratio circular tokamak, we thus have $\hat C = 1$, while in a stellarator with no bootstrap current, $\hat C=0$.
	
We use the recently implemented capability of SPEC to prescribe the toroidal current profile \citep{Baillod2021}, with the profiles defined in Eqs.(\ref{eq.volume current}) and (\ref{eq.bootstrapmodel}).
Unless stated otherwise, \ab{the Fourier resolution used in all results presented in this paper is $|n|\leq N=12$, $m\leq M=12$}, with $n$ the toroidal mode number and $m$ the poloidal mode number, meaning that \ab{$2[N+M(2N+1)]+1=625$} Fourier modes are used to describe each interface geometry. Results presented in this paper have been checked for convergence with respect to Fourier resolution \ab{(see appendix \ref{app.nvol dep})}.

In summary, we can construct free-boundary SPEC equilibria with a simple bootstrap current model and we are left with two free parameters, namely (i) $\beta$ which controls the total pressure in the system and (ii) $\hat{C}$, a dimensionless parameter, that controls the bootstrap current strength for a given plasma $\beta$.

\subsection{Scans over $\hat{C}$ and $\beta$}
\begin{figure}
	\centering
	\begin{tikzpicture}
		\node[] (p11) at (-3,2) {
			\includegraphics[width=.42\linewidth]{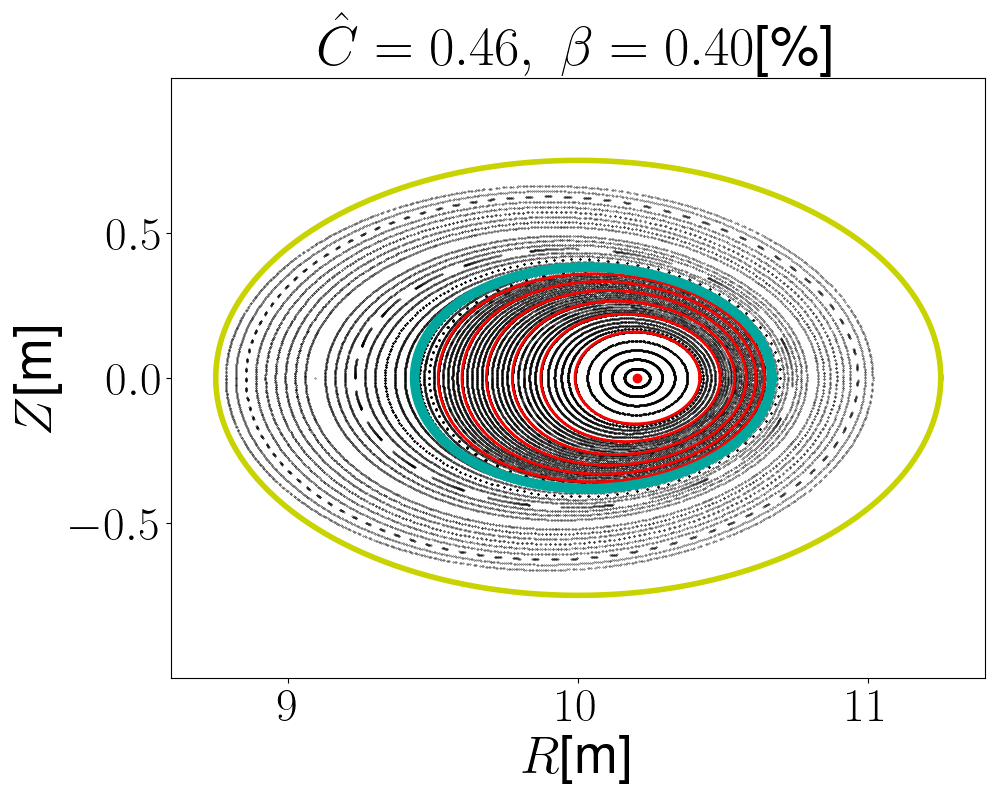}	
		};
		\node (p21) at (-3,-2.6) {
			\includegraphics[width=.42\linewidth]{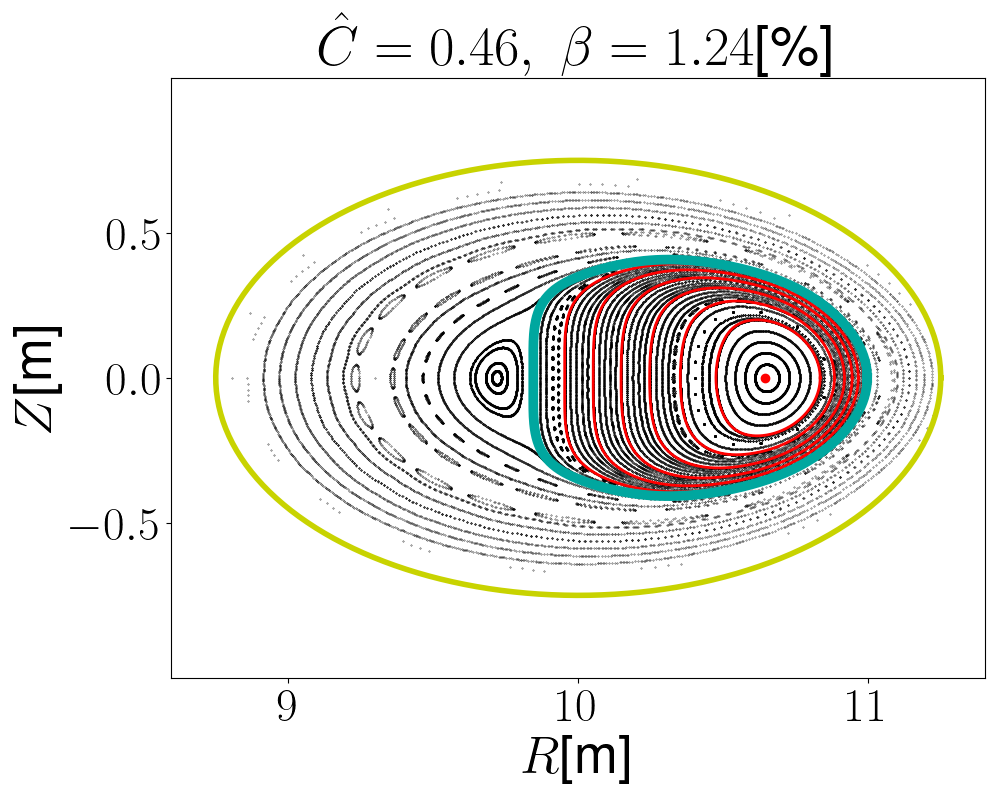}	
		};
		\node[] (p11) at (-3,-7.2) {
			\includegraphics[width=.42\linewidth]{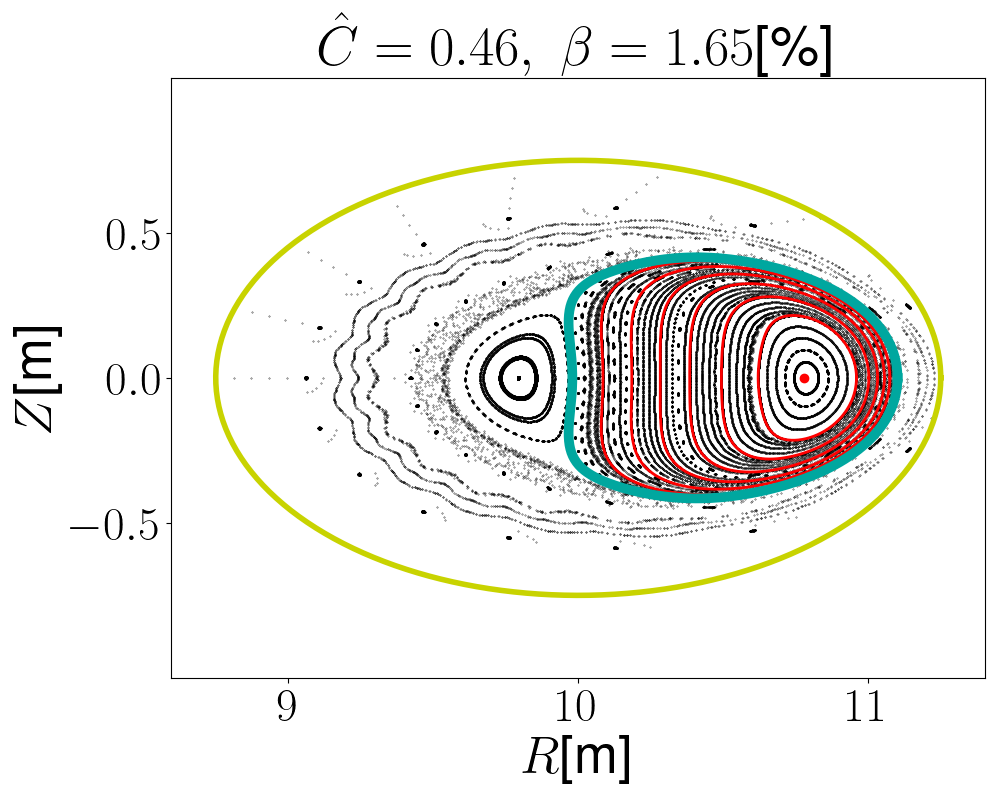}	
		};
		\node (p12) at (3,2) {
			\includegraphics[width=.42\linewidth]{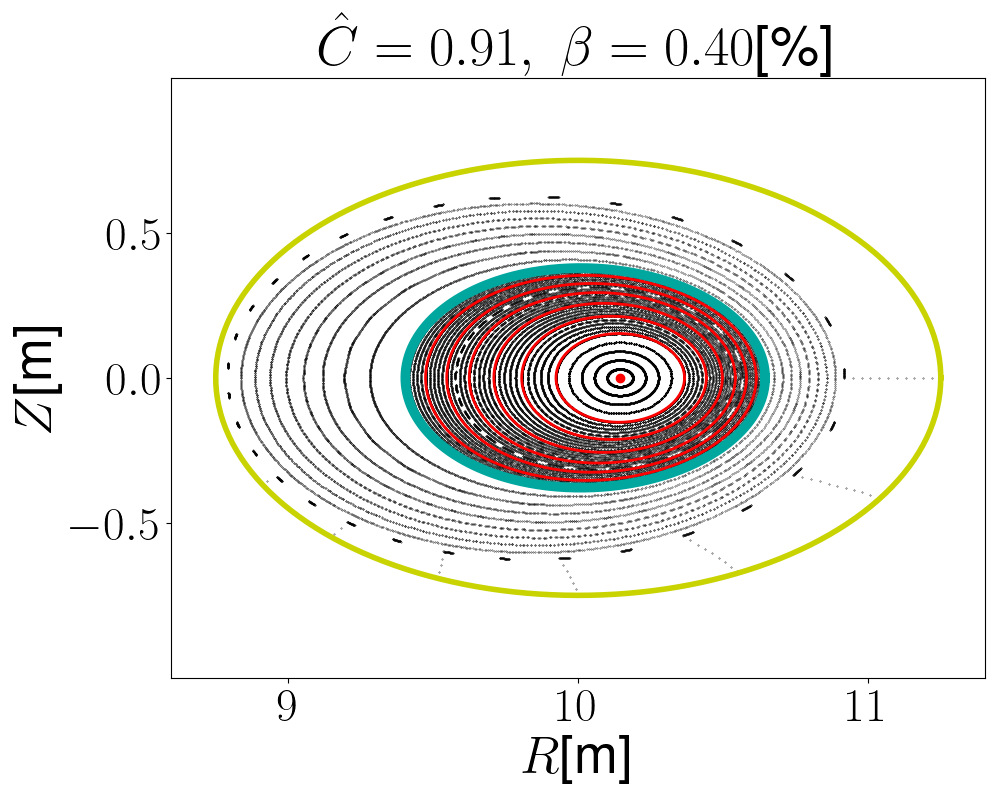}	
		};
		\node (p22) at (3,-2.6) {
			\includegraphics[width=.42\linewidth]{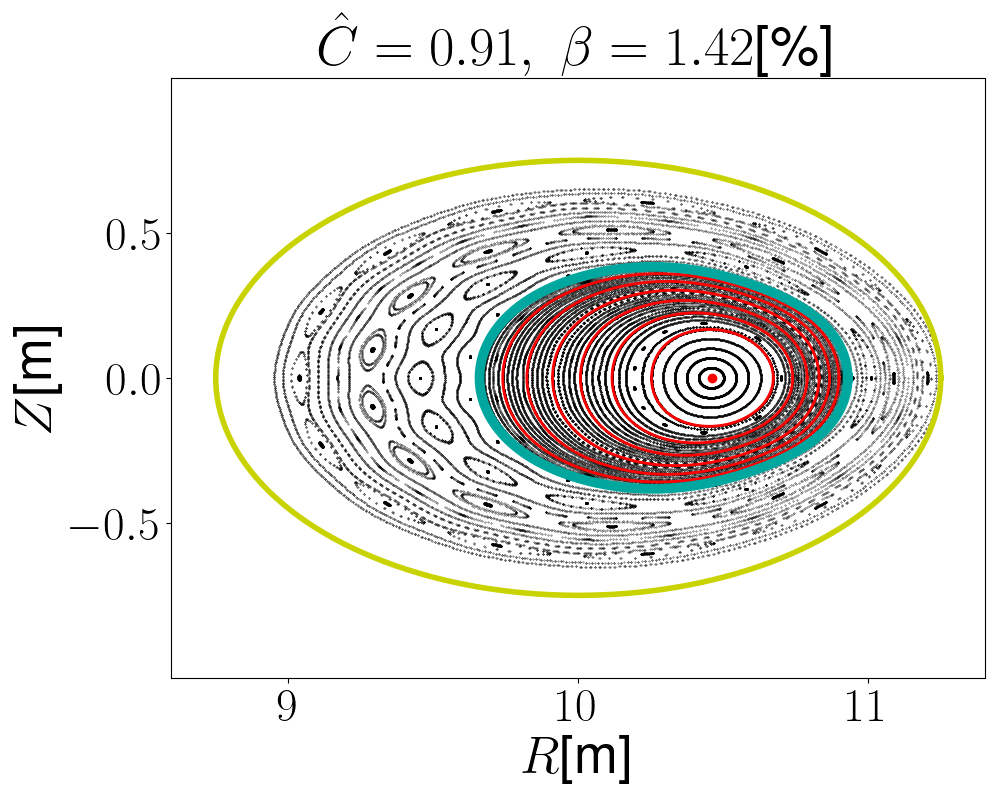}	
		};
		\node (p21) at (3,-7.2) {
			\includegraphics[width=.42\linewidth]{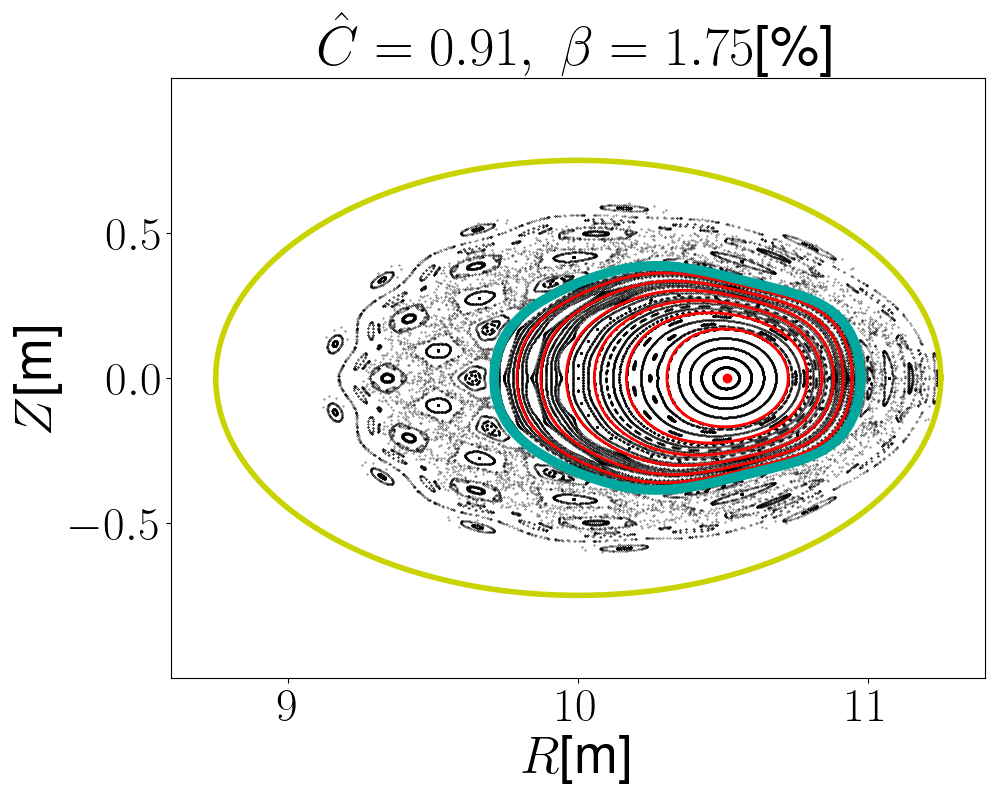}	
		};
		\draw[-stealth, very thick] (-6,3)--(-6,-7.7) node [midway, above, rotate=90] {Increasing $\beta$};
		\draw[-stealth, very thick] (-5,4.5)--(5,4.5) node [midway, above] {Increasing $\hat{C}$};	
	\end{tikzpicture}
	\caption{Poincar\'e plot (black dots) of equilibria at toroidal angle $\phi=0$ and at different values of $(\beta,\hat{C})$. Red lines: inner plasma volume interfaces; blue line: plasma boundary; and yellow line: computational boundary.  Left: $\hat{C}=0.46$. Right: $\hat{C}=0.91$.}
	\label{fig. poincare}
\end{figure}

\begin{figure}
	\centering
	\includegraphics[width=\linewidth]{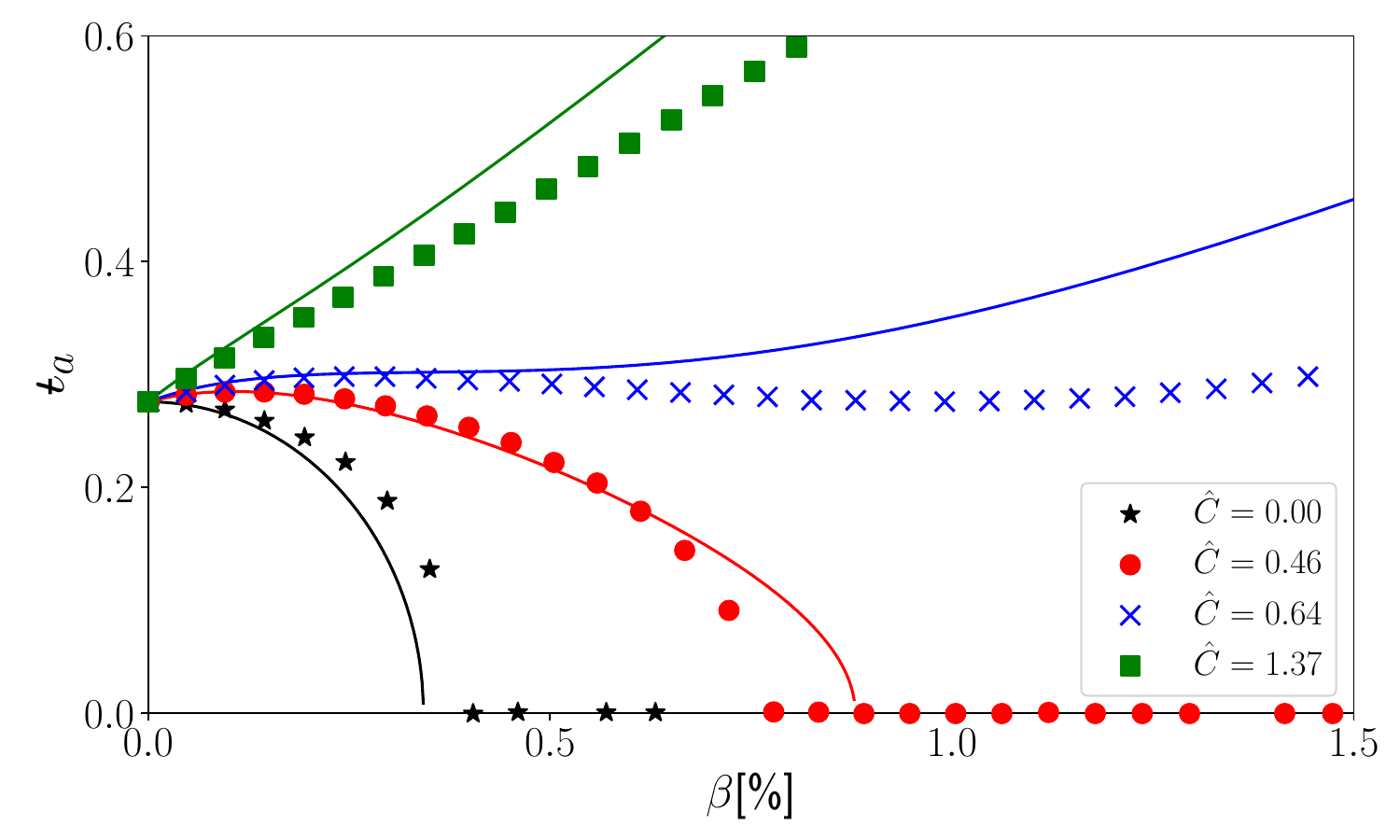}
	\caption{Edge rotational transform, $\iotabar_a$, as a function of plasma average $\beta$, for different values of $\hat{C}$; stars, circles, crosses and squares are SPEC calculations while full lines are given by Eq.(\ref{eq.iota_hbs}).}
	\label{fig. iota edge}
\end{figure}

A scan has been performed with $\beta\in[0,2\%]$ and $\hat{C}\in[0,2.26]$ representing $680$ SPEC calculations, each requiring about $24$ CPU-hours on the MARCONI cluster\footnote{https://www.hpc.cineca.it/hardware/marconi}. Figure~\ref{fig. poincare} shows some selected Poincar\'e sections at different values of $\beta$ and $\hat{C}$, while Figure~\ref{fig. iota edge} shows the edge rotational transform, \textit{i.e.} the rotational transform on the outer side of $\Gamma_{PB}$, as a function of $\beta$ for four different values of $\hat{C}$.

\begin{figure}
	\centering
	\includegraphics[width=\linewidth]{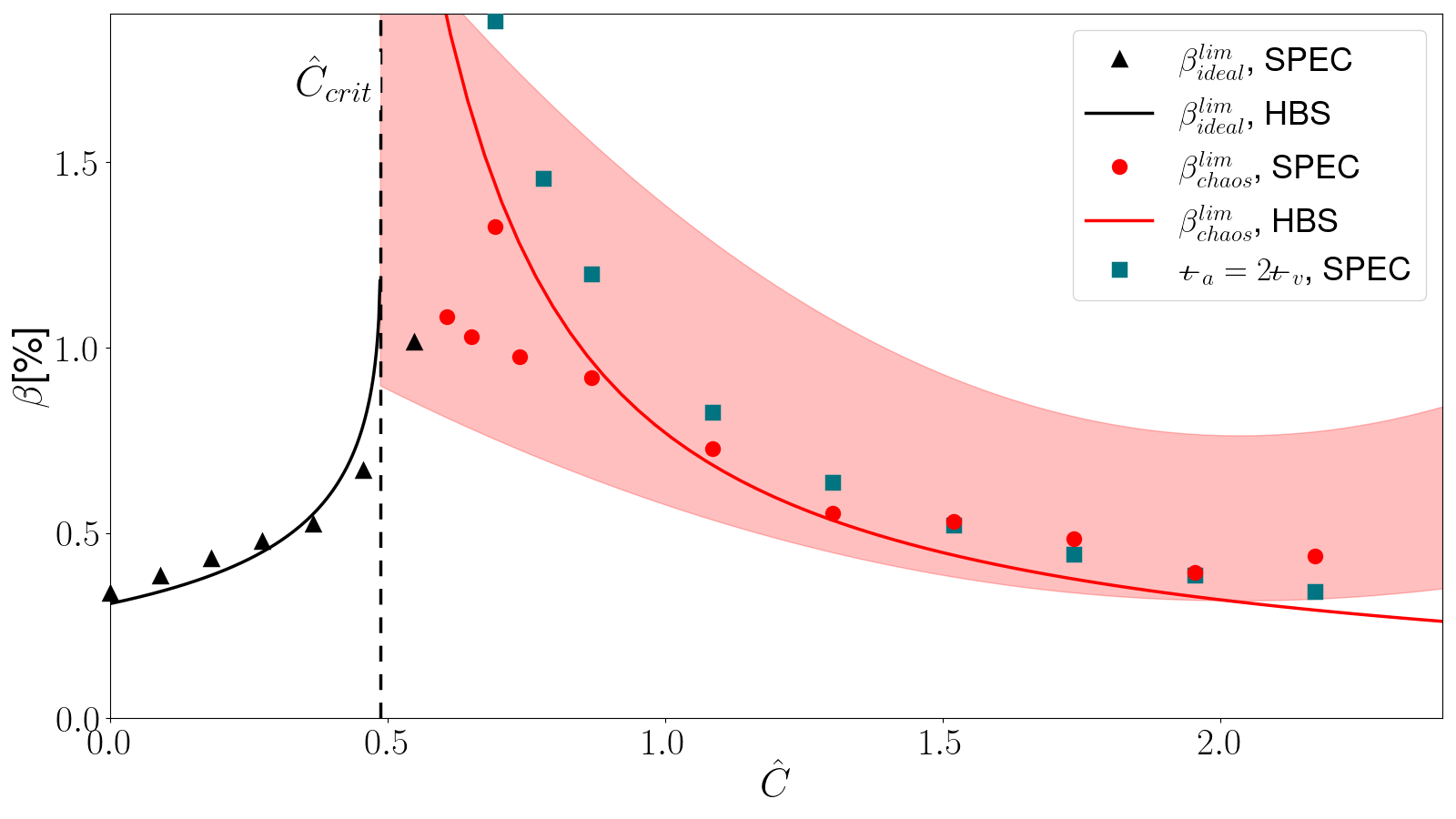}
	\caption{\ab{Equilibrium $\beta$-limit as a function of $\hat{C}$. Black triangles: ideal equilibrium $\beta$-limit, $\beta_{lim}^{ideal}$, as obtained from SPEC. Solid black line: analytical prediction for $\beta_{lim}^{ideal}$ from Eq.(\ref{eq.hbs ideal limit}). The dashed vertical line indicates the analytical value of $\hat{C}_{crit}$ from Eq.(\ref{eq.Ccrit analytical}). Red dots: values of the chaotic equilibrium $\beta$-limit, $\beta_{lim}^{chaos}$, obtained from SPEC for $ B_{r,crit}/B=10^{-5}$. The red area shows the range obtained from SPEC for $ B_{r,crit}/B\in[10^{-6},10^{-4}]$. Solid red line: analytical prediction for $\beta_{lim}^{chaos}$ obtained by solving Equation (\ref{eq.hbs chaos limit}). Blue squares: SPEC values for which $\iotabar_a=2\iotabar_v$.}}
	\label{fig. beta limits}
\end{figure}

For small values of $\hat{C}$, namely for $\hat{C} < \hat{C}_{crit}\approx 0.59$, the edge rotational transform decreases with increasing $\beta$ and eventually reaches zero (Figure~\ref{fig. iota edge}, black stars and red dots), at which point an $m=1,\ n=0$ island opens and forms a separatrix at the plasma boundary (see left panels of Figure~\ref{fig. poincare}). We will refer to this $\beta$-limit as the \emph{ideal equilibrium $\beta$-limit}, denoted by $\beta_{lim}^{ideal}$, since it is well described by ideal MHD theory (see section~\ref{sec. ideal limit}). The value of $\beta^{ideal}_{lim}$ obtained with SPEC is shown as a function of $\hat{C}$ in Figure~\ref{fig. beta limits} (\ab{black} triangles).

The ideal equilibrium $\beta$-limit can also be observed in tokamaks, although the underlying mechanism is different. In a tokamak, the plasma may be kept centered by applying a vertical magnetic field $B_Z$. As $\beta$ grows, $B_Z$ has to be increased, until it compensates the poloidal field $\mathbf{B}_p$ on the high field side. When this happens, the field is purely toroidal and a separatrix opens.
In a stellarator, the poloidal magnetic field does not have to cancel everywhere for a separatrix to open, it merely has to be such that a field line never completes a poloidal turn. If this happens, the edge rotational transform is zero and a separatrix opens.
In our calculations, the net toroidal current is constrained in the plasma volumes and at the interfaces. However the actual dependencies of the current density on the toroidal and poloidal angle are unconstrained. Pfirsch-Schl\"uter and diamagnetic currents angular dependencies are the source of the poloidal magnetic field perturbation, the lowering of the edge rotational transform, and ultimately the opening of the separatrix. This is why, even in a zero net-toroidal-current stellarator ($\hat{C}=0$), the edge rotational transform reaches zero.

For values of $\hat{C}>\hat{C}_{crit}$, the (now strong enough) bootstrap current is able to prevent the edge rotational transform from reaching zero for any $\beta$, and hence no $m=1,\ n=0$ island appears anywhere (see the blue crosses and green squares in Figure~\ref{fig. iota edge}). 
Instead, the edge rotational transform increases until many island chains open in the plasma and in the vacuum region (right panels of Figure~\ref{fig. poincare}). When these islands are large enough to have a significant impact on the radial transport, the \emph{chaotic equilibrium $\beta$-limit} is reached, denoted by $\beta_{lim}^{chaos}$. Finally, for all values of $\hat{C}$, islands start to overlap and generate large regions of chaotic field lines at sufficiently large values of $\beta$ (bottom panels of Figure (\ref{fig. poincare})).
In section~\ref{sec.diag}, a diagnostic to measure the critical value of $\beta$ at which the chaotic equilibrium $\beta$-limit is reached will be presented, and an analytical model that explain the results will be derived in section~\ref{sec. chaos limit}.

It may be argued that volume interfaces might not be able to support the pressure if islands or chaos are close by (see, for example, the bottom right panel in Fig.\ref{fig. poincare}) --- \textit{i.e.} that SPEC equilibria might not be trusted at large $\beta$ without further analyses. This question has been thoroughly studied in slab geometry by \citet{Qu2021}. They identified two reasons why a solution might not exist.


The first possibility is that the magnetic surface does not exist, in particular that it is fractal. In our calculations above the equilibrium $\beta$-limit, large magnetic islands and chaotic regions develop close to volumes interfaces. In this situation, it is indeed not known if the solution exists and additional analyses would be required, for example with convergence studies as proposed by \citet{Qu2021}.  Below the equilibrium $\beta$-limit, however, only small islands are present. The interfaces are not perturbed by neighbouring, large magnetic islands, and it is likely that the volume interfaces are magnetic surfaces. Since we are only interested in computing the equilibrium $\beta$-limit, it is sufficient to calculate equilibria \emph{below or equal to} the equilibrium $\beta$-limit; larger $\beta$ equilibria are irrelevant, and thus the question of existence of interfaces is eluded. \ab{In practice, we observe that large magnetic islands and chaotic field lines get close to the volume interfaces only for equilibria with $\beta$ sufficiently large to trust the results presented in this paper. Nevertheless, convergence studies have been performed, and results presented in this paper have been shown to be spectrally converged (see appendix \ref{app.nvol dep}).}

The second possibility is that the pressure jump on an interface is too large and a solution to the force-balance equation (\ref{eq. force balance}) does not exist. This is a possible explanation for when SPEC does not find an interface geometry that satisfies the force balance equation, Eq.(\ref{eq. force balance}). However, in our calculations, SPEC finds magnetic geometries that do satisfy force balance. This means that the pressure jump across the interfaces is small enough and a solution exists. To summarize this discussion, we can trust the SPEC solutions presented in this paper.

\section{Measure of magnetic chaos and its effect on radial transport}
\label{sec.diag}

\subsection{Fractal dimension, volume of chaos}
One approach to discriminate between a chaotic field line and other magnetic field line topologies is to evaluate the fractal dimension $D$ of the field line Poincar\'e section, for example using a box-counting algorithm \citep{Meiss1992c}. An almost binary behavior is then observed: either a magnetic field line stays on a magnetic surface whose Poincar\'e section is a one-dimensional object, $D=1$, or the magnetic field line has a fractal dimension $D>D_{crit}$, with $1<D_{crit}<2$. In our case, we observe that $D_{crit}=1.3$ can be used to differentiate between magnetic surfaces and chaos.
\citet{Loizu2017} proposed to evaluate the volume occupied by chaotic field lines with
\begin{equation}
	V_{chaos} = V_{total} \sum_{i=1}^{N_{lines}} \frac{(\psi_{t,i}-\psi_{t,i-1})}{\psi_a}\mathcal{H}(D_i-D_{crit}), \label{eq.volume chaos}
\end{equation}
where $N_{lines}$ is the number of considered field lines, $D_i$ is the fractal dimension of the $i^{\text{th}}$ line, $\mathcal{H}$ is the Heaviside function, $V_{total}$ is the total plasma volume, and $\psi_{t,i}-\psi_{t,i-1}$ measures the enclosed toroidal flux between field lines $i$ and $i-1$.

\begin{figure}
	\centering
	\includegraphics[width=0.75\linewidth]{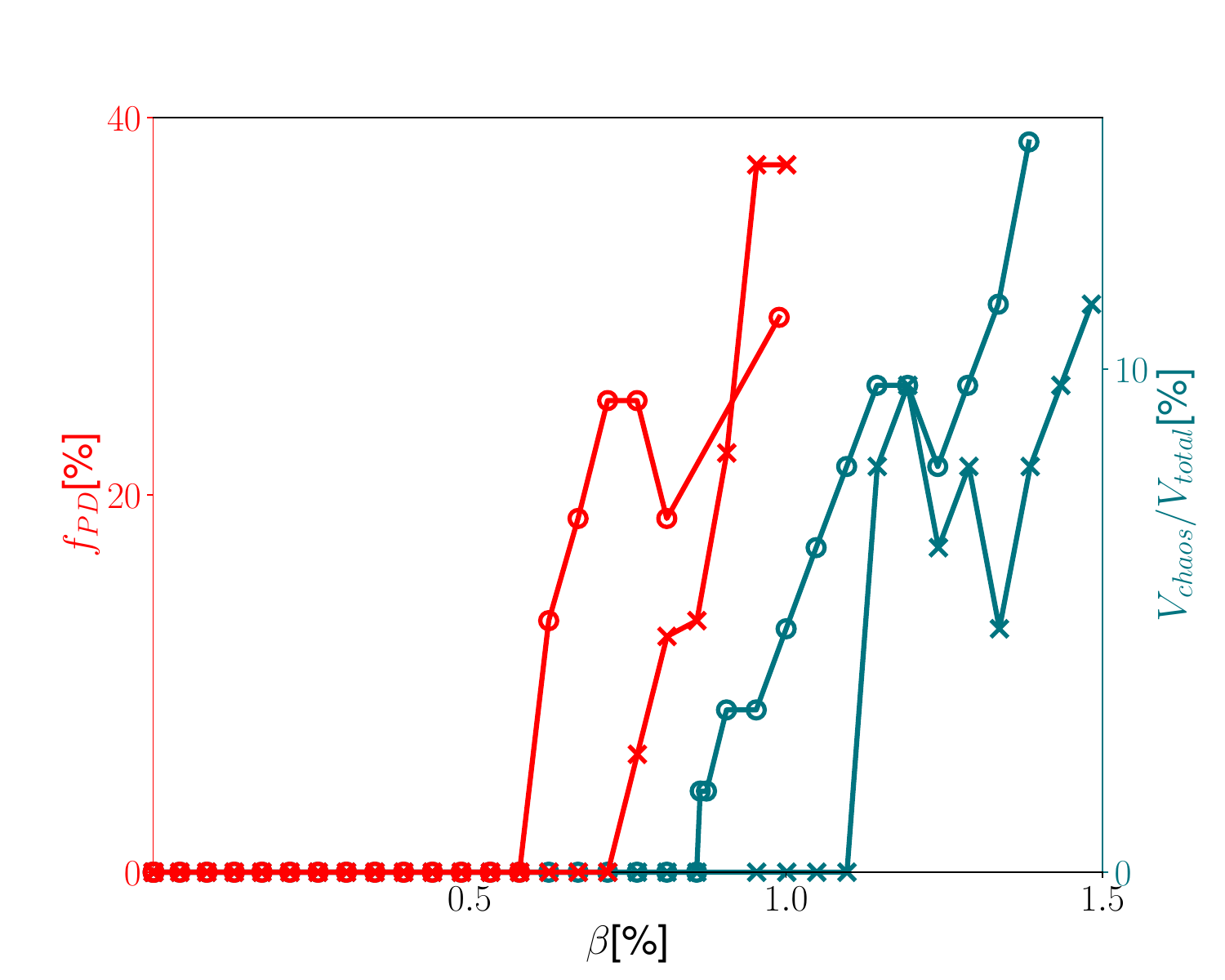}
	\caption{$V_{chaos}/V_{total}$ (blue), and $f_{PD}$ evaluated for $ B_{r,crit}/B=10^{-5}$ (red) versus plasma averaged $\beta$, for $M=N=8$ (crosses) and $M=N=10$ (circles).}
	\label{fig. metrics convergence}
\end{figure}

The chaotic equilibrium $\beta$-limit could then be defined as the $\beta$ above which $V_{chaos}>0$.
The volume of chaos, however, while very useful as a measure of the amount of chaotic field lines, does not provide enough information about whether or not the radial transport is enhanced by the destruction of magnetic surfaces. In addition, the volume of chaos is sensitive to the numerical resolution of the equilibrium --- the larger the number of Fourier modes, the greater the number of potential resonances in the equilibrium. Due to overlap between small islands chains generated by high order rationals, chaos may emerge at smaller $\beta$ as the Fourier resolution is increased. For example, in Figure~\ref{fig. metrics convergence} the volume of chaos is plotted as a function of $\beta$ for two different Fourier resolutions, $M=N=8$ and $M=N=10$ (blue lines). We see that with this diagnostic, the measured chaotic equilibrium $\beta$-limit would drop from $\sim 1.5\%$ to $\sim 1\%$ if it were defined as the $\beta$ above which $V_{chaos}>0$. However, in the $M=N=10$ scan, some of the chaotic field lines are formed by high order rationals and their associated smaller islands are expected to participate weakly to the radial transport, and could potentially be ignored. An alternative diagnostic that takes into account the effect of the magnetic field lines topology on the radial transport is thus required.

\subsection{Fraction of effective parallel diffusion}\label{sec.fpd}
We discuss an alternative measure to the volume of chaos to determine if the destruction of magnetic surfaces significantly impacts the radial transport. Here the parallel and perpendicular direction are defined as the direction along and across the magnetic field respectively, and the radial  direction $r$ as the direction perpendicular to isotherms, $\nabla T\times\nabla r=0$, with $T$ the temperature. In recent work, \citet{paulHeatConductionIrregular2022} discussed the properties of the anisotropic heat diffusion equation, $\nabla\cdot(\kappa_\parallel T + \kappa_\perp T)=0$, where $\kappa_\parallel$ and $\kappa_\perp$ are the parallel and perpendicular heat conductivities. In particular, \citeauthor{paulHeatConductionIrregular2022} demonstrated that, under the assumption that $\boldsymbol{\kappa}$ and $\nabla\cdot\boldsymbol{\kappa}$ are analytical, isotherms are topologically constrained to be toroidal surfaces --- this forbids isotherms to align with the magnetic field in regions occupied by magnetic islands and magnetic field line chaos. \ab{Here $\boldsymbol{\kappa}$ is the diffusion tensor, defined as $\boldsymbol{\kappa}=\kappa_\perp\mathbf{I}+(\kappa_\parallel-\kappa_\perp)\mathbf{B}\mathbf{B}/B^2$, with $\mathbf{I}$ the identity tensor.} Motivated by comparing the local parallel diffusion to the local perpendicular diffusion, \citeauthor{paulHeatConductionIrregular2022} introduced the \emph{volume of effective parallel diffusion}, which is the volume of plasma where the parallel heat transport dominates perpendicular heat transport,
\begin{equation}
V_{PD} = \frac{1}{V_{total}} \int_{V_{total}}\mathcal{H}(\kappa_\parallel|\nabla_\parallel T|^2-\kappa_\perp|\nabla_\perp T|^2) d\mathbf{x}^3, \label{eq.paul_volume_effective_diffusion}
\end{equation}
where the parallel and perpendicular gradients are defined as $\nabla_\parallel=\mathbf{B}(\mathbf{B}\cdot\nabla) / B^2$ and $\nabla_\perp=\nabla-\nabla_\parallel$ respectively. In regions occupied by magnetic islands and magnetic field line chaos, the constraint on the isotherms topology implies that the magnetic field has a non-zero radial component, thus $\nabla_\parallel T> 0$. Depending on the ratio $\kappa_\parallel / \kappa_\perp$, the volume of effective parallel diffusion can then be greater than zero. On the contrary, in regions occupied by magnetic surfaces, isotherms largely coincide with magnetic surfaces, which means that $\nabla_\parallel T$ is negligible, and consequently the volume $\mathcal{V}_{PD}$ is zero. Leveraging these properties, we can define the \ab{chaotic} equilibrium $\beta$-limit as the $\beta$ above which $V_{PD}>0$.

To determine the \ab{chaotic} equilibrium $\beta$-limit, it is only required to determine if $V_{PD}$ is zero or not; its absolute value is irrelevant. We thus construct a proxy function for $V_{PD}$ that does not depend on the temperature profile, but only on the magnetic field. We start by noticing that the Heaviside function in Eq.(\ref{eq.paul_volume_effective_diffusion}) is greater than zero when $\kappa_\parallel|\nabla_\parallel T|^2\ge \kappa_\perp|\nabla_\perp T |^2$. As we expect the radial magnetic field to be small in comparison to the total magnetic field, $B_r\ll B$, we can write $\nabla_\parallel T\sim \nabla T\ B_r /B$, and $\nabla_\perp T\sim \nabla T$. The volume of effective parallel diffusion is then greater than zero if there is a finite volume where
\begin{equation}
	\left(\frac{B_r}{B}\right)^2  \ge  \frac{\kappa_\perp}{\kappa_\parallel} \equiv \left(\frac{B_{r,crit}}{B}\right)^2.\label{eq.bcrit_estimate}
\end{equation}
Considering the electron heat transport as a figure of merit for the confinement properties of the equilibrium, and using the Spitzer-H\"arm conductivity for $\kappa_{\parallel,e}$ \citep{s.i.braginskiiTransportProcessesPlasma1965}, we get
\begin{equation}
	\left(\frac{ B_{r,crit}}{B}\right)^2 =  5.2\cdot 10^{-22}\frac{n_e\log\Lambda \  \chi_{\perp,e}}{T_e^{5/2}},
\end{equation}
where $\log\Lambda$ is the Coulomb logarithm, and typically $\chi_{\perp,e}=\kappa_{\perp,e}/n_e\sim 1 \text{m}^2\text{s}^{-1}$. Here everything is to be expressed in SI units except $T_e$, which is in $eV$. For temperatures and densities between $1$ to $10$ keV and $10^{19}$ to $10^{20}\text{m}^{-3}$ respectively, $ B_{r,crit}/B$ ranges from $10^{-6}$ to $10^{-4}$. For example using typical values for W7-X high performance experiments \citep{klingerOverviewFirstWendelstein2019}, \textit{i.e.} $n_e=4\cdot10^{19}\text{m}^{-3}$, $T_e=5\ \text{keV}$, we obtain a critical normalized radial magnetic field of $ B_{r,crit}/B\sim 10^{-5}$. As a side note, we remark that the criterion (\ref{eq.bcrit_estimate}) can also be derived by considering the radial heat diffusion equation for electrons, $q_r = -\kappa_{\perp,e} (\nabla_\perp T_e)_r - \kappa_{\parallel,e} (\nabla_\parallel T_e)_r \sim -\kappa_{\perp,e} dT_e/dr -\kappa_{\parallel,e} dT_e/dr B_r^2/B^2$. Magnetic islands and chaos play then an important role in setting the local heat radial transport when the second term on the right hand side of the heat diffusion equation is larger than the first one, which occurs when $B_r^2/B^2\ge \kappa_\perp/\kappa_\parallel$, recovering equation (\ref{eq.bcrit_estimate}).

The volume of effective parallel diffusion can thus be written using the criterion (\ref{eq.bcrit_estimate}),
\begin{equation}
	V_{PD} \sim \frac{1}{V_{total}} \int_{V_{total}}\mathcal{H}\left(\left[\frac{ B_{r}}{B}\right]^2-\left[\frac{ B_{r,crit}}{B}\right]^2\right) d\mathbf{x}^3.
\end{equation}
This measure is however unpractical for the purpose of this paper, as it would require to evaluate the radial magnetic field everywhere in the plasma. Instead, the radial magnetic field is evaluated where it is expected to be the largest, \textit{i.e.} on a selected number of rational surfaces. We then construct \emph{the fraction of parallel diffusion},
\begin{equation}
	f_{PD} = \frac{1}{N_{res}}\sum_{i=1}^{N_{res}} \mathcal{H}\left(\left[\frac{B_{r}}{B}\right]^2-\left[\frac{B_{r,crit}}{B}\right]^2\right), \label{eq.def metric}
\end{equation}
where $N_{res}$ is the number of considered resonances, and the algorithm used to evaluate the radial magnetic field $B_r$ from SPEC equilibria is described in section~\ref{sec. measure b r}. The fraction of effective parallel diffusion is then the fraction of resonances in the plasma that contribute to the transport, \textit{i.e.} the fraction of resonances over which the diffusion due to parallel dynamics dominates. Note that $f_{PD}\neq V_{PD}$, but if $f_{PD}=0$, we can expect $\mathcal{V}_{PD}$ to be zero, and the opposite is true as well. The fraction of parallel diffusion can then be used as a proxy function to determine if the volume of parallel diffusion is zero or not. The chaotic equilibrium $\beta$-limit, $\beta^{chaos}_{lim}$, is obtained by taking the value of $\beta$ above which $f_{PD}>0$ (see Figure \ref{fig. metrics convergence}). 

Note that this does not define an equilibrium $\beta$-limit from an experimental point of view --- the metric $f_{PD}$ is positive as soon as one resonance satisfies Eq.(\ref{eq.bcrit_estimate}), which would, in practice, only flatten the temperature and density profiles locally. It is certainly possible to increase the plasma averaged $\beta$ further by increasing the input power. Our metric $f_{PD}$ however informs us that the effect of field line topology starts to become important and has to be taken into account in transport calculations for $\beta>\beta^{chaos}_{lim}$. One could imagine to combine the volume of chaos given by Eq.(\ref{eq.volume chaos}) with the criterion given by Eq.(\ref{eq.bcrit_estimate}), and only consider resonances that span a sufficiently large volume \emph{and} that contribute significantly to the radial transport. This idea will not be explored in this paper, and is left for future studies.	

In practice, the metric $f_{PD}$ is greater than zero when relatively small islands in comparison to the plasma minor radius emerge (using $ B_{r,crit}/B=10^{-5}$). Thus, as long as the SPEC volumes are large enough to allow these islands to grow, the number of volumes does not affect the metric evaluation. In addition, given a sufficiently large number of volumes, the pressure profile is well resolved by the stepped-pressure approximation and thus the equilibrium does not depend strongly on the number of volumes (appendix \ref{app.nvol dep}).


\subsection{Measure of the radial magnetic field} \label{sec. measure b r}
\begin{figure}
	\centering
	\includegraphics[width=.75\linewidth]{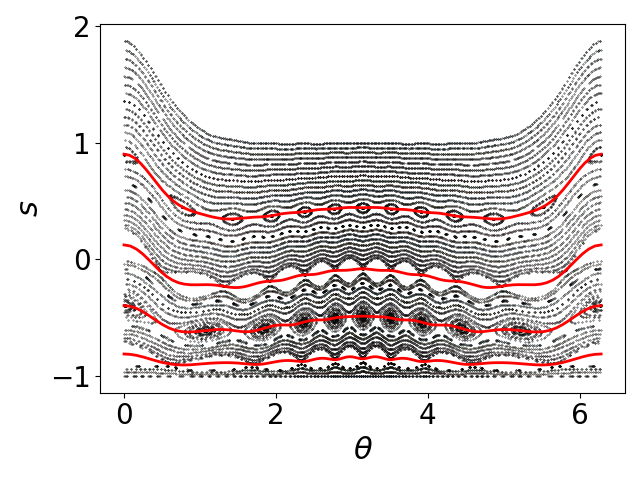}
	\caption{Black: Poincar\'e plot with magnetic surfaces and magnetic islands. Red: QFM surface $r=\text{const}$. The coordinate $s$ is a radial-like coordinate.}
	\label{fig.qfms_example}
\end{figure}

To evaluate the radial \ab{magnetic} field $ B_r$, it is useful to construct a general set of coordinates, such as quadratic flux minimizing (QFM) surfaces \citep{dewarAlmostInvariantManifolds1994, hudsonAlmostinvariantSurfacesMagnetic1996, hudsonConstructionIntegrableField1998}, or ghost surfaces \citep{hudsonAreGhostSurfaces2009}, which have been shown to coincide with isotherms \citep{Hudson2008}. We construct QFM surfaces  using the pyoculus package\footnote{\url{https://github.com/zhisong/pyoculus}}. These surfaces, thereafter named $\Gamma_{mn}$, are smooth toroidal surfaces that pass through the X- and O- points of the island chain corresponding to the $\iotabar=n/m$ rational resonant surface, and are constructed by finding the surfaces $\Gamma_{mn}$ minimizing the weighted quadratic flux $\int_{\Gamma_{mn}} w (B\cdot\mathbf{n})^2 dS$, where the weight $w$ is cleverly chosen such that the underlying Euler-Lagrange equation has non-singular solutions. Some examples of QFM surfaces are plotted in Figure~\ref{fig.qfms_example}. The radial coordinate $r$ is then defined as the direction perpendicular to the QFM surfaces.

We can now measure the radial component of the magnetic field at each resonant surface $\iotabar=n/m$. We start by identifying all potential resonances $(m,n)\in\mathbb{N}$ in each volume $\mathcal{V}_l$ \ab{within the plasma boundary}, such that (i) $n/m$ is within the rotational transform extrema in the volume, and (ii) $n$ is a multiple of the number of field periods. We construct QFM surfaces $\Gamma_{mn}$ for each of the identified resonances $\iotabar=n/m$. The magnetic field perpendicular to the QFM surface, $B_r$, is obtained by projecting  the magnetic field on their normal direction, and the magnetic field resonant harmonic, $B_{r,mn}$ is obtained after a standard Fourier transform of $ B_r$. \ab{Here, the poloidal angle is the straight-field line angle of the magnetic field tangential to the QFM surface. The Fourier spectrum of $B_r$ is largely dominated by the $(m,n)$ harmonic --- it has been verified that $B_{r,mn}$ is at least twice as large as the other Fourier harmonics of the radial magnetic field. We can thus assume $ B_r\approx B_{r,mn}$ to filter out numerical noise that may be generated by the QFM surface construction.}

Only resonances with large radial magnetic field will significantly participate to the radial transport. Since the magnetic field harmonics $B_{mn}$ are expected to decrease exponentially with the square of their mode numbers $m$ and $n$, \textit{i.e.} $B_{mn}\sim exp(-m^2-n^2)$, we can discard resonances with large poloidal and toroidal mode number and study only harmonics with mode number smaller than a given resolution, $m\leq M_{res}$ and $n\leq N_{res}$. In this paper, we set $M_{res}=25$ and $N_{res}=10$. 

With the definition of the \ab{chaotic} equilibrium $\beta$-limit from the fraction of effective parallel diffusion (\ref{eq.def metric}), only resonances with large radial magnetic field component matter; increasing the Fourier resolution of the equilibrium only introduces resonances with small radial magnetic field components, and thus \ab{has only a small impact on} the value of $f_{PD}$  --- see for example the comparison between two $\beta$-scans with resolution $M=N=8$ and $M=N=10$ in Figure~\ref{fig. metrics convergence} (red curves)\ab{, and the chaotic equilibrium $\beta$-limit convergence study in appendix \ref{app.nvol dep}}. Indeed, the critical $\beta$ at which $f_{PD}$ becomes larger than zero, namely $\beta_{lim}^{chaos}$, becomes quite insensitive to the Fourier resolution for sufficiently large values of $M$ and $N$, as the ones used for this paper ($M=N=12$). In that sense, this new diagnostic is more robust than the diagnostic based on the volume of chaos.

The chaotic equilibrium $\beta$-limit obtained using the metric $f_{PD}$ defined in Eq.(\ref{eq.def metric}) is plotted in Figure (\ref{fig. beta limits}) with \ab{a red shaded area}, spanning the range of $\beta_{lim}^{chaos}$ obtained when varying $ B_{r,crit}/B$ from $10^{-6}$ to $10^{-4}$. The value of $\beta^{chaos}_{lim}$ obtained for $ B_{r,crit}/B = 10^{-5}$ is shown with \ab{red} dots. We observe that the largest $\beta$-limit occurs at $\hat{C}\approx0.75$. A small, but non-zero bootstrap current thus \emph{increases} the equilibrium $\beta$-limit with respect to a classical stellarator without any net toroidal current ($\hat{C}=0$), and is thus beneficial.

\section{Analytical prediction for the equilibrium $\beta$-limits}\label{sec.analyticalmodel}


We now derive an analytical model that predicts both the ideal and chaotic equilibrium $\beta$-limits. We make use of high-$\beta$ stellarator expansion theories derived by \citet{wakataniStellaratorHeliotronDevices1998,Freidberg2014} to describe how the rotational transform at the plasma edge $\iotabar_a$ evolves with $\beta$, taking into account the effect of the bootstrap current as well. Once a formula for $\iotabar_a(\beta)$ has been derived, we can find whether an ideal $\beta$-limit is reached by solving $\iotabar_a(\beta)=0$. When no solution is possible, a chaotic $\beta$-limit may also be estimated by assuming that the edge iota is modified by order one with respect to the vacuum rotational transform, $\iotabar_a(\beta)-\iotabar_a(0) \sim \iotabar_a(0)$, at which point it is likely that many resonances exist. 

Assuming that (i) $\epsilon\ll1$, $\delta=|\mathbf{B}_p|/B_\phi\sim\epsilon^{3/4}$ with $\mathbf{B}_p$ the poloidal magnetic field, $\beta\sim\epsilon$ and $N_{fp}\sim\epsilon^{-1/2}$, that (ii) magnetic surfaces are circular, and (iii) considering Solove'v profiles for the pressure $dp/d\psi_p=\text{const}$, and the surface averaged toroidal current density $\langle j_\phi\rangle=\text{const}$, one can derive \citep{wakataniStellaratorHeliotronDevices1998,Freidberg2014} an analytical model for the edge rotational transform,
\begin{align}
	\iotabar_a &= (\iotabar_I+\iotabar_v)\sqrt{1-\nu^2}\label{eq.iota_hbs}\\ 
	\text{with}\ \iotabar_I &= 
	\frac{R_0}{2\psi_a}\mu_0I_\phi(\beta)\\ \label{eq.iota i}
	\text{and}\ \nu &= \frac{\beta}{\epsilon_a(\iotabar_I+\iotabar_v)^2},
\end{align}
where $I_\phi$ is the net toroidal current enclosed by the plasma and $\iotabar_v$ is the edge rotational transform in vacuum.

The bootstrap current model we employed in our equilibrium calculations (Eq.(\ref{eq.bootstrapmodel})) implies a linear relation between the net toroidal current in the system and the plasma $\beta$, thus
\begin{equation}
	\iotabar_I = \sigma\beta,
\end{equation}
where $\sigma$ is a proportionality constant. It can be related to $C$ by integrating Eq.(\ref{eq.current density continuous}) to compute $I_\phi$ in Eq.(\ref{eq.iota i}), leading to

\begin{equation}
	\sigma = \frac{2}{5}\frac{1}{\pi \epsilon_a^{3/2}\iotabar_v}\hat{C}.\label{eq.kappa-C}
\end{equation}

Combining Eqs.(\ref{eq.iota_hbs})-(\ref{eq.kappa-C}), analytical expressions of the edge rotational transform as a function of $\beta$ for different values of $\hat{C}$ can be obtained. Figure~\ref{fig. iota edge} compares the analytical curves to results obtained with SPEC.  We observe reasonable agreement especially at low $\beta$. As $\beta$ increases however, Eq.(\ref{eq.iota_hbs}) consistently underestimates the actual value of the rotational transform found by SPEC. Thus, even though the equilibrium constructed in section~\ref{sec.equil} does not exactly satisfy the assumptions used to derive Eq.(\ref{eq.iota_hbs}), the assumptions are reasonable enough to use this analytical model to understand our numerical results. Equation (\ref{eq.iota_hbs}) provides indeed an analytical (non-linear) relation for $\iotabar_a(\beta)$ which can be used to predict both the ideal and chaotic $\beta$-limits, as described in the following subsections.

\subsection{Ideal equilibrium $\beta$-limit \label{sec. ideal limit}}
The solution to the relation $\iotabar_a(\beta^{ideal}_{lim})=0$ is given by
\begin{align}
	\beta^{ideal}_{lim} = \frac{1}{\epsilon_a\sigma^2}\left[\frac{1}{2}-\iotabar_v\epsilon_a\sigma - \sqrt{1-4\iotabar_v\epsilon_a\sigma}\right], \label{eq.hbs ideal limit}
\end{align}
which is real for $\sigma<(4\iotabar_v\epsilon_a)^{-1}$, or 
\begin{equation}
	\hat{C} \leq \frac{5}{8}\frac{\psi_a}{\epsilon_a^{3/2}R_0^2B_0} \equiv \hat{C}_{crit}. \label{eq.Ccrit analytical}
\end{equation}
Note the limit
\begin{equation}
	\lim_{\sigma\rightarrow 0}\beta^{ideal}_{lim} = \epsilon_a\iotabar_v^2, \label{eq. beta limit C=0}
\end{equation}
retrieving the result from \citet{Freidberg2014} and \citet{Loizu2017} for a zero-net-current stellarator ($\hat C=0$).

The curve $\beta^{ideal}_{lim}(\hat{C})$ is plotted in Figure~\ref{fig. beta limits} with a black line. We observe that as $\hat{C}$ increases, the ideal equilibrium $\beta$-limit increases. Comparison with data points measured from SPEC equilibria (red triangles) shows good agreement, especially for weaker bootstrap current ($\hat C<0.5$). The analytical value of $\hat{C}_{crit}\approx0.48$ is reasonably close to the one obtained with SPEC (smaller by about $18\%$).


\subsection{Chaotic equilibrium $\beta$-limit \label{sec. chaos limit}}
For larger values of $\hat C$, \textit{i.e.} $\hat{C}>\hat{C}_{crit}$, the \ab{chaotic} equilibrium $\beta$-limit is due to the emergence of chaos and its effectiveness in increasing the transport, thus estimating the chaotic equilibrium $\beta$-limit with Eq.(\ref{eq.iota_hbs}) is not trivial - it is not known, \textit{a priori}, which resonance will participate to the radial transport first. However it is reasonable to assume that when the bootstrap current modifies the edge rotational transform by order one with respect to $\iotabar_v$, \textit{i.e.}
\begin{equation}
	\Delta\iotabar_a\equiv\iotabar_a-\iotabar_v=\iotabar_v, \label{eq.condition chaos}
\end{equation}
magnetic islands and chaos are expected to appear. The values of $\beta$ computed with SPEC at which the condition Eq.(\ref{eq.condition chaos}) is satisfied are plotted with brown squares in Figure~\ref{fig. beta limits}. We observe good agreement with the chaotic equilibrium $\beta$-limit (blue dots) for $\hat{C}>1$.
	
We can also directly solve equation (\ref{eq.condition chaos}) using equation (\ref{eq.iota_hbs}). We obtain a fourth order polynomial equation for $\beta$,
\begin{equation}
	\beta^4 + 4\frac{\iotabar_v}{\sigma}\beta^3 + \left(2\frac{\iotabar_v^2}{\sigma^2}-\frac{1}{\epsilon_a^2\sigma^4}\right)\beta^2 - 4\frac{\iotabar_v^3}{\sigma^3}\beta - 3\left(\frac{\iotabar_v}{\sigma}\right)^4 = 0. \label{eq.hbs chaos limit}
\end{equation}

The real, positive root of Eq.(\ref{eq.hbs chaos limit}) is plotted with \ab{a red line} in Figure~\ref{fig. beta limits}. Direct comparison with the numerical data (blue squares) shows that Eq.(\ref{eq.hbs chaos limit}) consistently underestimates the values of $\beta$ that satisfy Eq.(\ref{eq.condition chaos}); this is a direct consequence of the underestimate of $\iotabar_a$ by the analytical model (Figure~\ref{fig. iota edge}). The general dependence on $\hat{C}$ is however recovered, capturing the chaotic equilibrium $\beta$-limit trend (red dots in Figure~\ref{fig. beta limits}) observed numerically for values of \ab{$\hat{C}>1$. We remark that there are no free parameters in this analytical model. For $\hat{C}_{critical}<\hat{C}<1$, the analytical model (\ref{eq.hbs chaos limit}) overestimates greatly the chaotic equilibrium $\beta$-limit obtained with SPEC. In this transition region, the edge rotational transform depends weakly on $\beta$ for $\beta\lesssim 1$ (see, for example, the blue crosses in Figure~\ref{fig. iota edge}). As a consequence, the solution to Eq.(\ref{eq.condition chaos}) is large, and is therefore a bad estimate for the chaotic equilibrium $\beta$-limit. A more refined model would be required to better reproduce the results.}


\subsection{Dependence on design parameters}
The edge rotational transform in vacuum is approximately equal to the rotational transform on axis (low shear configuration), and can be estimated by a zeroth order near axis expansion \citep{helanderTheoryPlasmaConfinement2014,Loizu2017},
\begin{equation}
	\iotabar^{axis}_v\approx\iotabar_v = \frac{N_{fp}}{2}\frac{(r_{max}-r_{min})^2}{r_{max}^2+r_{min}^2}.
\end{equation}

\begin{figure}
	\centering
	\includegraphics[width=\linewidth]{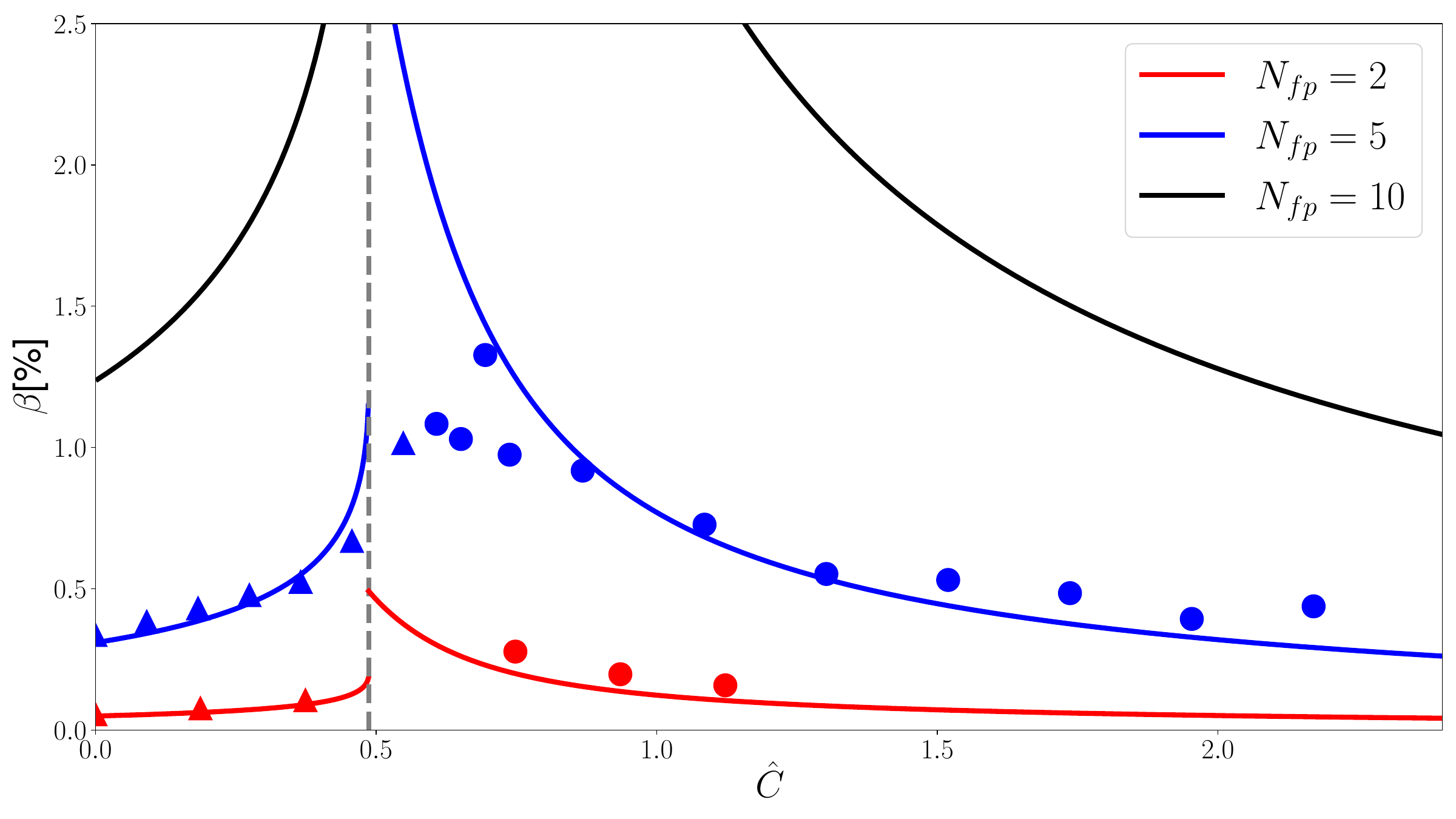}
	\caption{Analytical predictions of the equilibrium $\beta$-limit for different numbers of field period $N_{fp}$. Full lines: ideal and chaotic equilibrium $\beta$-limits as predicted by Eq.(\ref{eq.hbs ideal limit}) and  Eq.(\ref{eq.hbs chaos limit}) respectively. Triangles and dots: ideal and chaotic equilibrium $\beta$-limits as obtained by SPEC respectively.}
	\label{fig. analytical curves}
\end{figure}

For low values of $\hat{C}$, the ideal equilibrium $\beta$-limit grows with the vacuum rotational transform (see equation (\ref{eq. beta limit C=0})). For example, increasing the number of field periods increases $\iotabar_v$, thus also the equilibrium $\beta$-limit, as shown in Figure~\ref{fig. analytical curves}. These results were corroborated by SPEC calculations with $N_{fp}=2$, \ab{while calculations with $N_{fp}=10$ were difficult to achieve due to SPEC numerical fragility issues}.

More generally, any mechanism that increases the rotational transform in vacuum will increase the ideal and chaotic equilibrium $\beta$-limits. An increase in rotational transform can be achieved by either increasing the number of field periods, increasing the ellipse eccentricity (\textit{i.e.} increasing the harmonic $R_{11}=Z_{11}$) or adding some torsion to the magnetic axis. Magnetic axis torsion can however have a strong impact on the computed equilibrium, and additional studies would be required to see if it affects the conclusions of this paper.

Equation (\ref{eq.Ccrit analytical}) gives $\hat{C}_{crit}=0.48$, \textit{i.e.} the equilibrium $\beta$-limit is maximized for a bootstrap current that has half the strength of the bootstrap current in an equivalent circular tokamak. Interestingly, if we approximate the total toroidal flux in the plasma as $\psi_a\approx\pi a_{eff}^2 B_0$, \ab{with $B_0$ the modulus of the magnetic field on axis}, we get $\hat{C}_{crit}=5\ab{\pi} \sqrt{\epsilon_a}/8$, which only depends on the inverse aspect ratio.

\section{Conclusion}
\label{sec.conclusion}
The SPEC code has been used to perform a large number of free-boundary stellarator equilibrium calculations including bootstrap current that allowed us to completely characterize classical stellarators in terms of their equilibrium $\beta$-limit. For configurations with low bootstrap current ($\hat{C}<\hat{C}_{crit}$), an ideal equilibrium $\beta$-limit has been identified, where a central $(m,n)=(1,0)$ island appears. 
Stronger bootstrap current ($\hat{C}>\hat{C}_{crit}$) prevents this central island to open. Instead, a chaotic equilibrium $\beta$-limit is reached, where the radial heat transport generated by pressure-induced magnetic islands and magnetic field line chaos competes with turbulence. 
We have implemented a proxy function to determine if the effective volume of parallel diffusion proposed by \citet{paulHeatConductionIrregular2022} is greater than zero, thereby assessing the impact of the field line topology on radial transport and deducing the equilibrium $\beta$-limit from SPEC equilibrium calculations.

An analytical model showed good agreement with the ideal equilibrium $\beta$-limit obtained numerically \ab{for weak bootstrap current}. The general trend for the chaotic equilibrium $\beta$-limit could also be extracted for \ab{stronger bootstrap current, up to $\hat{C}\sim 2$}. Analytical insights provided ways to predict the effect of design parameters on the equilibrium $\beta$-limit; for example, the ideal $\beta$-limit has been shown to increase with $\hat{C}$, while the chaotic equilibrium $\beta$-limit decreases with $\hat{C}$, thereby showing a peak equilibrium $\beta$-limit around $\hat{C}_{crit}$. The critical value of $\hat{C}_{crit}$ depends only on the inverse aspect ratio\ab{, under reasonable assumptions}.

To improve the equilibrium $\beta$-limit of stellarators, optimization of different parameters can be performed. For example, \citet{Landreman2021a} recently coupled SPEC with the simsopt framework \citep{Landreman2021b} to perform optimization for good magnetic surfaces at the same time as quasisymmetry in vacuum, and \citet{Baillod2022} showed that good magnetic surfaces can be recovered in finite $\beta$, finite current equilibria by modifying either the plasma boundary, the coils, or by injecting a toroidal current in the plasma. Applying the same recipe to a sequence of equilibria with increasing $\beta$, one can optimize a stellarator configuration for larger equilibrium $\beta$-limit. Note however that the fraction $f_{PD}$ is generally not a smooth function of the equilibrium and might not be a good target function for optimization. Another smooth function should be developped from the radial magnetic field component $ B_{r}$ if one desires to minimize the impact of field line topology on radial transport.

Future studies will focus on more exotic stellarator geometries, for example configurations optimized for quasisymmetry or quasi-isodynamicity, and include self-consistent bootstrap currents, as proposed by \citet{landremanOptimizationQuasisymmetricStellarators2022}. Finally, one could use the SPEC code to evaluate the stability limit for different values of $\hat C$, using the methods developed by \citet{Kumar2021,Kumar2022}. This would provide useful information on the dependence of the stability limit on the parameter $\hat C$, and allow comparison with the equilibrium $\beta$-limit.

\section{Acknowledgments} 
The authors thank P. Helander, S. R. Hudson, C. Zhu, J. Schilling, J. Cazabonne and E. Balkovic for useful discussions. This work was supported in part by the Swiss National Science Foundation. \ab{This work has been carried out within the framework of the EUROfusion Consortium, via the Euratom Research and Training Programme (Grant Agreement No 101052200 — EUROfusion) and funded by the Swiss State Secretariat for Education, Research and Innovation (SERI). Views and opinions expressed are however those of the author(s) only and do not necessarily reflect those of the European Union, the European Commission, or SERI. Neither the European Union nor the European Commission nor SERI can be held responsible for them.} This research was supported by a grant from the Simons Foundation (1013657, JL).

\section*{Data Availability Statement}

The data that support the findings of this study are available from the corresponding author upon reasonable request.

\section*{Declaration of interests}

The authors report no conflict of interest.

\appendix
\section{Convergence on numerical resolution and number of volumes}\label{app.nvol dep}
\begin{figure}
	\centering
	\includegraphics[width=.85\linewidth]{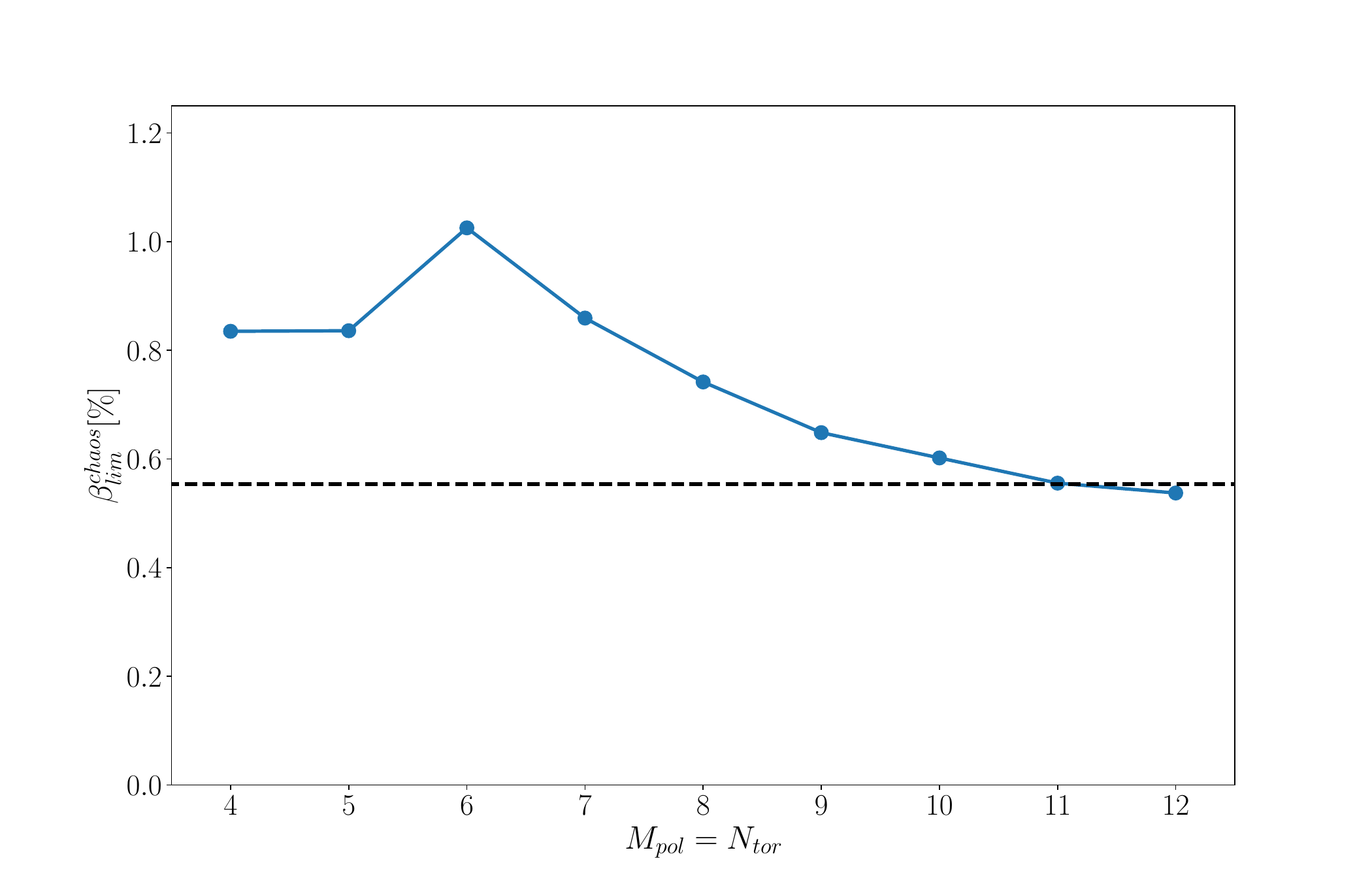}
	\caption{Chaotic equilibrium $\beta$-limit, as obtained by SPEC for $B_{r,crit}/B_0=10^{-6}$, and $\hat{C}=1.37$ as a function of the equilibrium Fourier resolution. The black, dashed line is the analytical prediction obtained by solving Eq.(\ref{eq.hbs chaos limit}).}
	\label{fig.fourier convergence}
\end{figure}

The chaotic equilibrium $\beta$-limit, as obtained by SPEC for $B_{r,crit}/B_0=10^{-6}$, and $\hat{C}=1.37$, is plotted on Figure \ref{fig.fourier convergence} as a function of the equilibrium Fourier resolution. Convergence towards a value close to the analytical prediction is observed. The resolution used to compute the results presented in this paper, \textit{i.e.} $M_{pol}=N_{tor}=12$, seems large enough so that the relative variations obtained by further increasing the Fourier resolution become small, of the order of $3\%$.

\begin{figure}
	\centering
	\includegraphics[width=.7\linewidth]{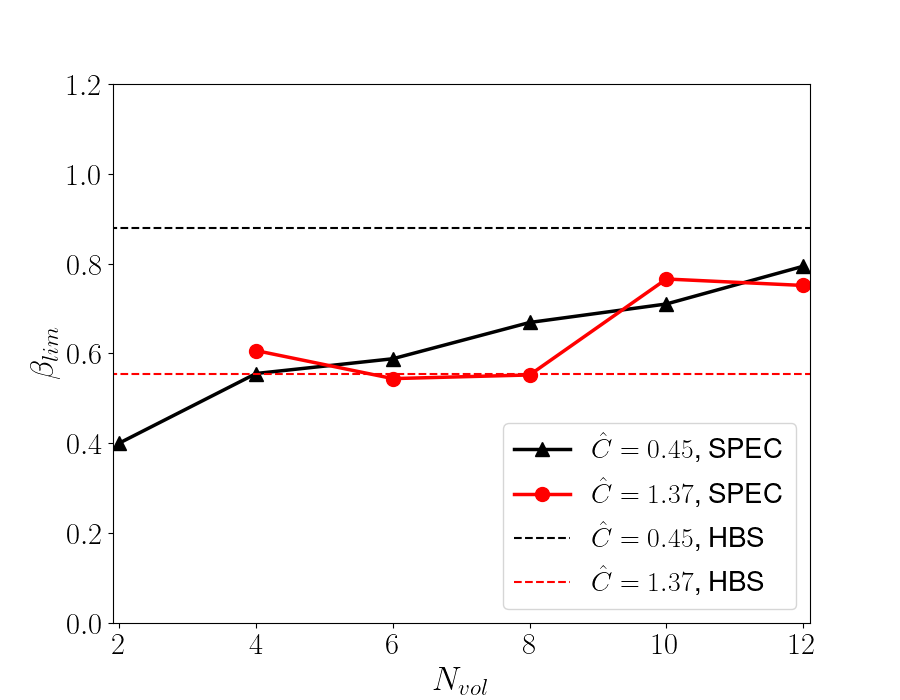}
	\caption{Dependence of the equilibrium $\beta$-limit on the number of volumes $N_{vol}$. Black: ideal equilibrium $\beta$-limit, obtained for $\hat{C}=0.45$. Red: chaotic equilibrium $\beta$-limit, obtained for $\hat{C}=1.37$. Full lines: equilibrium $\beta$-limit, as obtained with SPEC, and dashed line: analytical prediction, as obtained with the HBS theory.}
	\label{fig.nvol dependence}
\end{figure}

We now discuss the dependence of the ideal and the chaotic equilibrium $\beta$-limit dependence on the number of volumes $N_{vol}$. Keeping the same coil shapes and currents, we approximate the pressure profile $p=p_0(1-\psi_t/\psi_a)$ with different number of interfaces supporting an equal pressure step, $[[p]]_l = p_0 / N_{vol}$. We set $I^v_{\phi,l}=0$ and $I^s_{\phi,l}$ following the bootstrap current model described in Eq.(\ref{eq.bootstrapmodel}). A large range of $\beta$ is scanned for $\hat{C}=0.46<\hat{C}_{crit}$ and $\hat{C}=1.37>\hat{C}_{crit}$, and for $N_{vol}=\{2,4,6,8,10,12\}$. The corresponding ideal and chaotic equilibrium $\beta$-limits obtained from SPEC are shown on Figure \ref{fig.nvol dependence}.

For $\hat{C}=0.45$, an ideal equilibrium $\beta$-limit is found, that grows with the number of volumes, asymptotically approaching the analytical prediction as obtained by Eq.(\ref{eq.hbs ideal limit}). This is expected, as ideal MHD is recovered as the number of volumes approaches infinity \citep{Dennis2013}. Interestingly, we observe variations of the chaotic equilibrium $\beta$-limit of the order of $30\%$ as the number of volumes is changed, clearly within the range of values covered when $B_{r,crit}$ is varied. The dependence of the chaotic equilibrium $\beta$-limit is thus negligible in comparison to its dependence on $B_{r,crit}$, \textit{i.e.} on the plasma temperature and densities. We conclude that the stepped-pressure assumption made by the SPEC model has little to no consequences on the physical results presented in this paper.

\bibliographystyle{jpp}

\bibliography{references}

\end{document}